\documentclass[prl,aps,superscriptaddress,preprintnumbers,nofootinbib,nobibnotes]{revtex4}
\usepackage{graphicx}
\usepackage{amsfonts}
\usepackage{amssymb}
\usepackage{amsmath}
\usepackage{dsfont}
\usepackage{hyperref}
\usepackage{slashed}
\usepackage{empheq}
\usepackage{multirow}
\usepackage{fancyhdr}
\usepackage{appendix}
\usepackage{subfigure}
\usepackage{times}

\DeclareMathOperator{\hz}{h\relax{\kern-.15em}z}
\DeclareMathOperator{\pz}{\psi\relax{\kern-.15em}z}

\usepackage[english]{babel}
\usepackage{amssymb}
\usepackage{amsfonts}
\usepackage{amsmath}
\usepackage{graphicx}
\usepackage{epsfig}
\usepackage{psfrag}
\newcommand{\be}{\begin{equation}} \newcommand{\ee}{\end{equation}}
\newcommand{\bea}{\begin{eqnarray}} \newcommand{\eea}{\end{eqnarray}}
\newcommand{\beann}{\begin{eqnarray*}}  \newcommand{\eeann}{\end{eqnarray*}}
\newcommand{\bfig}{\begin{figure}} \newcommand{\efig}{\end{figure}}
\newcommand{\ba}{\begin{array}} \newcommand{\ea}{\end{array}}
\newcommand{\bcen}{\begin{center}} \newcommand{\ecen}{\end{center}}
\newcommand{\btab}{\begin{tabular}} \newcommand{\etab}{\end{tabular}}

\newcommand{\matt}{\left ( \begin{array}{ccc}}
    \newcommand{\ematt}{\end{array} \right )} \newcommand{\matf}{\left ( \begin{array}{cccc}}
    \newcommand{\ematf}{\end{array} \right )} \newcommand{\vect}{\left ( \begin{array}{c}}
    \newcommand{\evect}{\end{array} \right )}    \def\beqn{\begin{eqnarray}}
 \def\eeqn{\end{eqnarray}}  
     
   \renewcommand{\Im}{\mathop{\rm Im}}
\newtheorem{Proposition}{Proposition}[section]

\newtheorem{Theorem}{Theorem}[section]
\newtheorem{Lemma}{Lemma}[section]
\newtheorem{Corrolary}{Corrolary}[section]

\newcommand{\bp}{\begin{Proposition}}	\newcommand{\ep}{\end{Proposition}}
\newcommand{\bt}{\begin{Theorem}}	\newcommand{\et}{\end{Theorem}}
\newcommand{\bl}{\begin{Lemma}}		\newcommand{\el}{\end{Lemma}}
\newcommand{\bc}{\begin{Corrolary}}	\newcommand{\ec}{\end{Corrolary}}

\hypersetup{
colorlinks=true,
linkcolor=blue,
filecolor=magenta,
urlcolor=cyan,
citecolor=blue,
}

\begin{document}

\preprint{IFT-UAM/CSIC-24-101}

\title[Pseudospectra of Quasinormal Modes and Holography]{Pseudospectra of Quasinormal Modes and Holography} 

\author{Daniel Are{\'a}n}\email{daniel.arean@uam.es}
\affiliation{Departamento de F{\'i}sica Te{\'o}rica, Universidad Aut{\'o}noma de Madrid, Campus de Cantoblanco, 28049 Madrid, Spain}
\affiliation{Instituto de F{\'i}sica Te{\'o}rica UAM-CSIC, 
	C/ Nicol{\'a}s Cabrera 13-15, Campus de Cantoblanco, 28049 Madrid, Spain}

\author{ David Garcia-Fari{\~n}a}\email{david.garciafarinna@estudiante.uam.es}
\affiliation{Departamento de F{\'i}sica Te{\'o}rica, Universidad Aut{\'o}noma de Madrid, Campus de Cantoblanco, 28049 Madrid, Spain}
\affiliation{Instituto de F{\'i}sica Te{\'o}rica UAM-CSIC, 
	C/ Nicol{\'a}s Cabrera 13-15, Campus de Cantoblanco, 28049 Madrid, Spain}

\author{Karl Landsteiner}\email{karl.landsteiner@csic.es}
\affiliation{Instituto de F{\'i}sica Te{\'o}rica UAM-CSIC, 
	C/ Nicol{\'a}s Cabrera 13-15, Campus de Cantoblanco, 28049 Madrid, Spain}

\begin{abstract}
The holographic duality (also known as AdS/CFT correspondence or gauge/gravity duality) postulates that strongly coupled quantum field theories can be described in a dual way in asymptotically Anti-de Sitter space.
One of the cornerstones of this duality is the description of thermal states as black holes with asymptotically Anti-de Sitter boundary conditions. This idea has led to valuable insights into such fields as transport theory and relativistic hydrodynamics. In this context, the quasinormal modes of such black holes play a decisive role and therefore their stability properties are of upmost interest for the holographic duality. 
We review recent results using the method of pseudospectra. 
\end{abstract}

\pacs{ }

\maketitle

\setcounter{secnumdepth}{4}


\section{Introduction}
\subsection{Blitz review of the holographic duality}
Before discussing the role of quasinormal modes we need first to understand the basics of the AdS/CFT correspondence. 
Gauge/gravity duality has its roots in Maldacena's conjecture that Type IIB string theory on AdS$_5$ $\times$ S$_5$ is dual to ${\mathcal N}=4$ supersymmetric gauge theory \citep{Maldacena:1997re,Aharony:1999ti}.\footnote{Both descriptions arise from the low energy limit of a stack of $N$ D3-branes in string theory in flat 10-dimensional spacetime \citep{Aharony:1999ti}.} 
Let us quickly unpack this statement. ${\mathcal N}=4$ supersymmetric gauge theory is 
a non-abelian, four-dimensional quantum field theory whose field content consists of six scalars, four Majorana fermions and a gauge field. They all transform under the adjoint representation of the gauge group $SU(N)$.  It features four supersymmetries and this fixes all the couplings between the different fields. As it is a gauge theory physical observables are gauge invariant operators such as $\mathrm{tr}(F_{\mu\nu} F^{\mu\nu})$. The global symmetry group $SO(6)$ acts on the scalars and the fermions (in the $SU(4)$ spin representation of $SO(6)$). Besides the theory has conformal symmetry and 32 supercharges. 

The dual theory is a theory of gravity (type IIB string theory) but living in  ten dimensions.
Five of these are a geometric realization of the internal $SO(6)$ symmetry as the isometry of the five-dimensional sphere S$_5$. Supersymmetry is generated by two ten-dimensional spinors of equal chirality which also results in 32 supercharges. Conformal symmetry arises as the isometry group on AdS$_5$.

The field theory gauge coupling $g_{YM}$ and  rank of the gauge group $N$ are related to the dual string theory string coupling $g_s$ (the amplitude for a string to split in two) and to the ratio of AdS$_5$ curvature scale $R=-20/L$ and string scale $l_s$
in the following way:
\begin{eqnarray} \label{eq:holodictionary}
	g^2_{YM} N &\propto& \frac{L^4}{l_s^4}\,,\\
	1/N &\propto& g_s\,.
\end{eqnarray}
Gauge/gravity duality is therefore a strong weak coupling duality: for weak curvature we have large $L$ and therefore also large 't-Hooft coupling $g_{YM}^2N$. In this regime of weak curvature stringy effects are negligible and we can approximate the string theory by type IIB supergravity. If we furthermore take the rank of the gauge group $N$ to be very large we can also neglect quantum loop effects and end up with classical supergravity. 
This is the form of the correspondence most useful for the applications 
to many body physics: classical (super)gravity on $(d+1)$-dimensional Anti-de Sitter space is the infinite coupling and infinite rank limit of a gauge theory in $d$ dimensions. 

This is now promoted to a principle: (quantum-)gravity in asymptotically 
$(d+1)$-dimensional Anti-de Sitter space can be understood as a strong coupling version of a dual quantum field theory in $d$ dimensions \citep{Zaanen:2015oix,Ammon:2015wua}. 
For applications to quantum field theory the most useful coordinate system is the so-called Poincar\'e patch 
\begin{equation}\label{eq:ads}
	ds^2 = \frac{r^2}{L^2} ( -dt^2 + d\vec{x}^2 ) + \frac{L^2}{r^2}dr^2\,.
\end{equation}
The space on which the dual quantum field theory lives is recovered by taking the limit 
$ds^2_\mathrm{QFT} = \lim_{r\rightarrow \infty} r^{-2} ds^2$. 

Since the correspondence relates a $(d+1)$-dimensional theory to a $d$-dimensional one it is also called ``holographic'' duality.
The radial coordinate has a physical interpretation as energy scale. The high-energy or UV limit in the field theory is identified with the $r\rightarrow \infty$ limit in the AdS geometry whereas the low-energy IR limit is $r\rightarrow 0$. 

On shell the asymptotic behavior of the fields in AdS in a large $r$ expansion is
\begin{equation}\label{eq:asymptoticexpansion}
	\Phi(r,x) = r^{-\Delta_-} \left(\Phi_0(x) + O(r^{-2})\right) + r^{-\Delta_+} \left(\Phi_1(x) + O(r^{-2} )\right)\,.
\end{equation}
The exponents $\Delta_{\pm}$ obey $\Delta_- < \Delta_+$ and depend on the nature of the field, e.g. for a scalar field of 
mass $m$ they are $\Delta_{\pm} = \frac 1 2 (d\pm \sqrt{d^2+4m^2L^2})$. We note that for a scalar field in asymptotically AdS, masses in the range $-d^2/4<m^2<0$ are perfectly regular and do not imply any acausality or instability \citep{Breitenlohner:1982jf}.

It turns out that the leading solution given by $\Phi_0(x)$ is non-normalizable and thus non-dynamical. It is interpreted as a boundary condition $\Phi_0(x)=J(x)$ on the AdS field $\Phi(r,x)$. The classical on-shell action  becomes a functional of these boundary conditions $S_\mathrm{cl}[J]$. In the (super)gravity limit the on-shell action is interpreted as the generating functional of (connected) Green's functions $Z_\mathrm{c}[J]=S_\mathrm{cl}[J]$ in the dual field theory. The boundary condition $J$ is now interpreted as a source for an operator $\mathcal{O}$ in the dual field theory whose correlation functions can be obtained from
\begin{equation}\label{eq:defGF}
	\left\langle \mathcal{O}_1(x_1) \dots  \mathcal{O}_n(x_n) \right\rangle = \frac{\delta^n S_\mathrm{cl}}{\delta J_1(x_1) \dots \delta J_n(x_n)}\,.
\end{equation}
More specifically the expectation value of the operator $\mathcal{O}$ is given by
\begin{equation}
	\left\langle \mathcal{O}(x)\right\rangle \propto \Phi_1(x)\,.
\end{equation}
In this way the leading and subleading terms in the asymptotic expansion  \eqref{eq:asymptoticexpansion} have dual field theory interpretations. The mass range $-d^2/4\leq m^2\leq0$ corresponds to renormalizable operators.\footnote{We note that this is the so-called standard quantization scheme and allows only for operators of dimensions larger than $d/2$. In order to describe operators of smaller dimensions one needs to exchange the role of source and operator (``alternative quantization''). For further details on that see \cite{Klebanov:1999tb}.}

Generically the equation of motion for $\Phi(r,x)$ is a second order partial differential equation. In order to solve it one needs to supply additional boundary conditions. 
The metric \eqref{eq:ads} has a (degenerate) horizon at $r=0$ and it was argued in \citep{Son:2002sd} that for time dependent solutions retarded Green's functions of the dual quantum field theory,
\begin{equation}\label{eq:retarded}
G_R(t,\vec{x}) = -i \Theta(t)\left \langle [\mathcal{O}(t,\vec{x}), \mathcal{O}(0,0) ] \right \rangle\,,
\end{equation}
are obtained by imposing {\em infalling} boundary
conditions.

The infalling boundary condition is of course the main ingredient for the existence of quasinormal modes.
In Anti-de Sitter space it does however not lead to quasinormal modes due to the fact that the horizon is degenerate. The corresponding (holographic) retarded Green's function does not have poles but rather a branch cut along the positive real axis \cite{Son:2002sd}. 
This changes as soon as we consider a black hole with asymptotic AdS boundary conditions with planar horizon topology (AdS black brane). Its line element for $d=4$ is 
\begin{eqnarray}\label{eq:blackbrane}
ds^2&=& \frac{r^2}{L^2}\left( - f(r)dt^2 + d\vec{x}^2\right) + \frac{L^2 }{r^2f(r)}dr^2\,, \\
f(r) &=& 1- \frac{r_h^4}{r^4}\,.
\end{eqnarray}
This metric has a non-degenerate horizon at $r=r_h$. The Hawking temperature is $\pi T L^2=r_h$. The holographic (or gauge/gravity) interpretation is that the dual field theory is now in a thermal state with the temperature given by the Hawking temperature \cite{Gubser:1996de,Witten:1998zw}. 

The field $\Phi$ is expanded in (boundary) plane waves
\begin{equation}
\Phi(r,t,\vec{x}) = \int \frac{d\omega d^{3}k}{(2\pi)^4} \tilde\Phi_0(\omega, \vec k ) e^{-i\omega t + i \vec{k}\vec{x} }\, F_{\omega,\vec{k}}(r)\,.
\end{equation}
For every fixed $\omega, \vec{k}$ the linearized equation of motion for the fluctuation boils down then to an ordinary second order differential equation for $F_{\omega,\vec{k}}$. 
The point at infinity is a regular singular point with characteristic exponents $\Delta_\pm$. 
We impose infalling boundary conditions by demanding that $ F_{\omega,\vec{k}} 
\sim e^{-i\omega(t+r_*)}$ on the horizon (we use a tortoise coordinate here $dr_* = dr/f(r)$ such that the horizon sits at $r_*\rightarrow -\infty$). The asymptotic expansion of $F_{\omega,\vec{k}}(r)$ is 
\begin{equation}
F_{\omega,\vec{k}} = A(\omega, \vec{k})\, r^{-\Delta_-} \left[1 + O(1/r)\right] + B(\omega,\vec k )\, r^{-\Delta_+} \left[ 1+ O(1/r)\right] \,,
\end{equation}
where $A(\omega,\vec{k})$ and $B(\omega,\vec{k})$ are the Fourier transforms of $\Phi_0(x)$ and $\Phi_1(x)$ respectively.
The Fourier transform of the retarded two point Green's function  \eqref{eq:retarded} can now be calculated as 
\begin{equation}\label{eq:Grholographic}
\tilde{G}_R(\omega,\vec k ) = K \frac{B(\omega, \vec k )}{A(\omega, \vec k )} \,,
\end{equation}
where $K$ is some normalization constant \cite{Son:2002sd}.

\subsection{Holographic quasinormal modes}
The definition of the holographic retarded Green's function depends on a subtlety. It is impossible to calculate a retarded (or advanced) Green's function from an action as is indicated in eq. \eqref{eq:defGF}. In thermal field theory one needs to go to the Schwinger-Keldysh formalism in which the time coordinate lives on a (complex) contour \citep{Bellac:2011kqa}. It turns out that the Schwinger-Keldysh contour is naturally implemented on the maximally analytic extension of the AdS black brane metric. In that case one has a second boundary on which the direction of the timelike Killing vector is reversed in comparison to the one covered by the coordinate patch  \eqref{eq:blackbrane}. Strictly speaking retarded holographic Green's functions can only be {\it defined} on this maximally analytically continued double sided Kruskal type manifold \citep{Herzog:2002pc}. 
Infalling boundary conditions correspond then to analytic continuation of the solution to the whole Kruskal manifold. For all practical purposes the retarded Green's function can however be {\em computed} on the patch \eqref{eq:blackbrane} by the simple recipe \eqref{eq:Grholographic}. The quasinormal modes describe the return to the thermal equilibrium \citep{Horowitz:1999jd}. Their frequencies are the poles of the holographic Green's function \eqref{eq:Grholographic} in the complexified $\omega$ plane \citep{Birmingham:2001pj,Kovtun:2005ev}.

Retarded two point functions are the central objects in linear response theory. The response in the operator $\mathcal{O}$ under a perturbation (source) $J(t,x)$ with Fourier transform $\tilde J(\omega,\vec k)$ is
\begin{equation}
\left \langle \mathcal{O}(t,\vec{x})\right \rangle = \int \frac{d\omega d^3 k}{(2\pi)^4} \, e^{-i \omega t + i \vec{k}\vec{x}} 
G_R(\omega,\vec k ) \tilde J(\omega, \vec k ) = -i 
\Theta(t) \int \frac{d^3k}{(2\pi)^3}\sum_n R_n(k) \tilde J(\omega_n(\vec k), \vec k )  e^{-i\omega_n t + i\vec{k}\vec{x} }\,.
\end{equation}
where $R_n$ are the residues of $G_R$ at the poles $\omega_n$\footnote{We assume here that there are no contributions from the integral along the large radius half circle in the lower complex $\omega$ half-plane.}. As long as all quasinormal frequencies lie in the lower half of the complex $\omega$-plane the response decays exponentially fast. A mode in the upper half indicates an instability leading eventually to a phase transition.

A special role is played by linearized perturbations of gauge fields and the metric. In this case the dual operator corresponds to a conserved current and the quasinormal modes spectrum contains so-called hydrodynamic modes \citep{Policastro:2002se}, i.e. those fulfilling  
\begin{equation}\label{eq:hydromode}
\lim_{k\rightarrow 0} \omega_H(\vec k) = 0\,.
\end{equation}
For a gauge field  one finds in this way a diffusive mode that obeys in the small $|\vec{k}|$ limit 
$\omega_\mathrm{diffusive}= -i D \vec k^2$ where the diffusion constant $D=\frac{1}{2\pi T}$. The metric fluctuations contain a shear-channel with a similar diffusive law $\omega_\mathrm{shear} = - i \frac{\eta}{\epsilon + p} \vec k^2$ where $\epsilon + p = s T$ are the energy density $\epsilon$, pressure $p$ and entropy density $s$ of the dual field theory. Famously one finds $\frac{\eta}{s}=\frac{1}{4\pi}$ \citep{Kovtun:2004de}.

In some exceptional cases exact solutions for the holographic Green's function can be found. If there are only three regular singular points of the differential equation it can then be mapped to the hypergeometric differential equation. This happens for the case of a gauge field in the five-dimensional AdS black brane background at vanishing momentum $\vec k=0$. The holographic retarded Green's function is  \citep{Myers:2007we}
\begin{equation}
G_R(\omega) = K \left[ 2i\omega + 
\omega^2 \psi\left(\frac{(1-i)\omega}{4}\right)     + 
\omega^2 \psi\left(-\frac{(1+i)\omega}{4}\right) 
\right]\,,
\end{equation}
where $\psi(z)$ is the digamma function. The poles are at\footnote{We have rescaled the frequency such that the physical values are $\omega_\mathrm{phys}=\pi T \omega$. We further note that the surface gravity is $\kappa=2\pi T$.} $\omega_n = 2n\,(\pm 1 - i)$. More generally the corresponding differential equation has more than three regular singular points and cannot be solved exactly. In these cases one needs to resort to numerical approximations.

\section{Pseudospectra of holographic quasinormal modes}
The infalling boundary conditions on the horizon of the AdS black brane have the consequence that the differential operator is a non-Hermitian and non-normal one. Its eigenvalues are complex numbers, precisely the quasinormal frequencies. It is a well known fact that eigenvalues of non-normal operators suffer from spectral instability. This means that a small perturbation of the operator can change the value of the eigenvalues dramatically. In fact it is this spectral instability that makes the prediction and calculation of quasinormal frequencies challenging. The method of pseudospectra has emerged as an ideal tool to assess the spectral instability of non-normal operators in a quantitative (and also qualitative) way \citep{trefethen2000spectral}. 

\begin{figure}
	\centering
	\includegraphics[width=.35\linewidth]{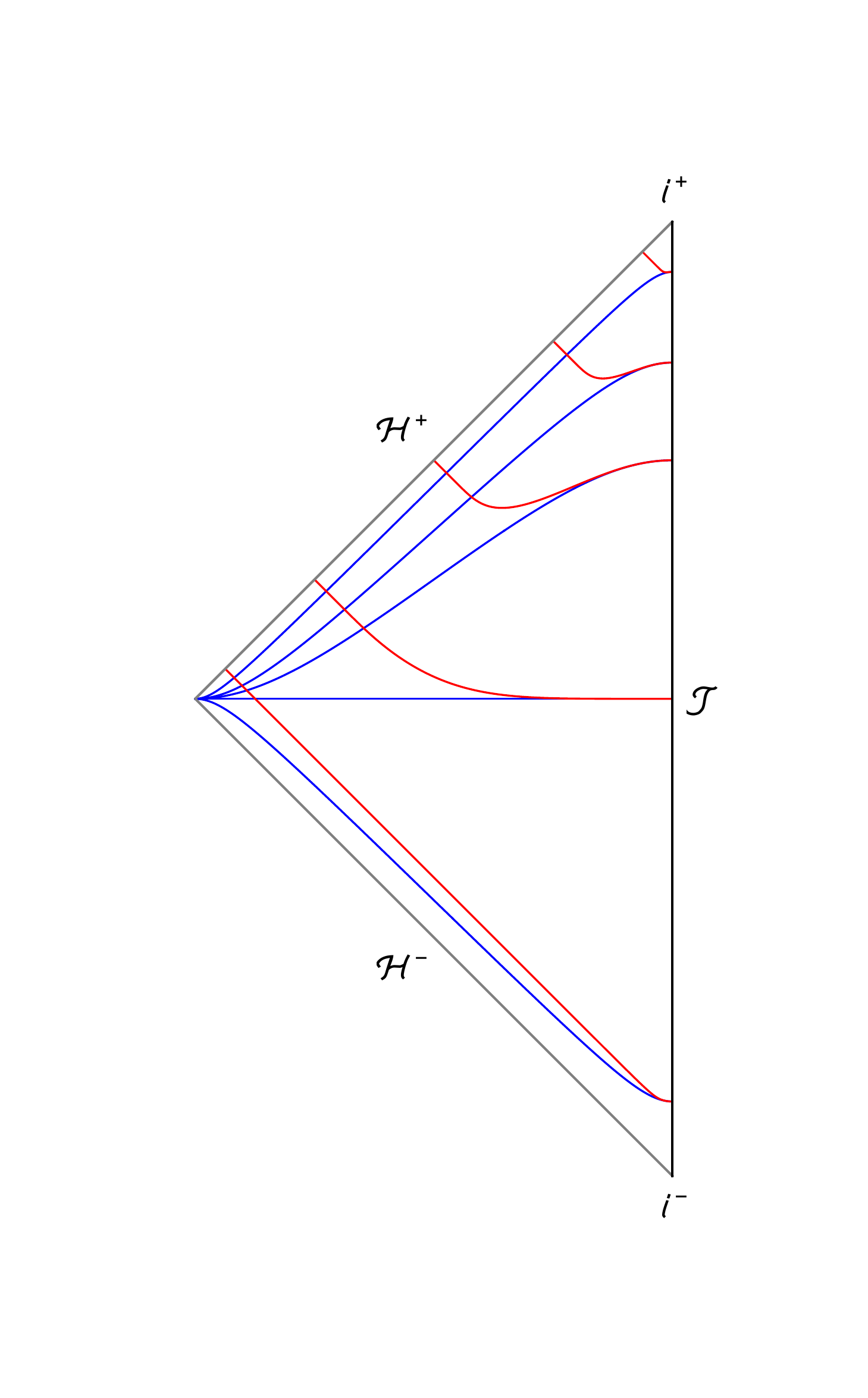}
	\caption{Penrose Diagram of the exterior region of SAdS$_{4+1}$. The AdS boundary is denoted by $\mathcal{J}$, $\mathcal{H}^{+}$ ($\mathcal{H}^{-}$) represents the future (past) horizon and $i^+$ ($i^-$) denotes the future (past) time-like infinity. The red lines correspond to constant $\tau$ hypersurfaces  \eqref{eq:RegularCoordinates} and the blue lines represent constant $t$ hypersurfaces \eqref{eq:blackbrane}. }
	\label{fig:PenroseDiagram_RegCoords}
\end{figure}

The calculation of pseudospectra of quasinormal modes was pioneered in \citep{Jaramillo:2020tuu} and further explored in \citep{Destounis:2021lum, Cheung:2021bol, Gasperin:2021kfv, Jaramillo:2021tmt,Jaramillo:2022kuv, Berti:2022xfj, Konoplya:2022pbc, alsheikh:tel-04116011,Destounis:2023ruj,Boyanov:2022ark,Cao:2024oud,Sarkar:2023rhp, Destounis:2023nmb,Arean:2023ejh, Cownden:2023dam, Boyanov:2023qqf,Courty:2023rxk}
in various astrophysical and cosmological contexts. We will concentrate here on the simple case of pseudospectra for a gauge field in the AdS black brane \citep{Arean:2023ejh}.
Pseudospectra answer the question of how far a quasinormal frequency can be displaced by a given perturbation of size $\epsilon$. This means of course that we need a way to measure the size of an operator that can be added as perturbation. Consequently we need to define an appropriate measure on a function space that contains the quasinormal modes. On physical grounds it is generally suggested to use a suitable norm based on the energy functional. Only in certain coordinate systems the quasinormal modes have ``nice'' or regular behaviour on the horizon. It turns out that in the coordinates of eq. \eqref{eq:blackbrane} the energy functional is not well defined. There are two strategies to deal with this problem. One is to use infalling Eddington-Finkelstein coordinates. These are often used in the literature on holographic quasinormal modes and the energy functional is indeed well-defined.
Another approach is to use so-called regular coordinates that interpolate between the Schwarzschild-type coordinates near the boundary and infalling Eddington-Finkelstein coordinates near the horizon \citep{Warnick:2013hba}. See figure \ref{fig:PenroseDiagram_RegCoords} for the Penrose diagram illustrating the geometrical nature of this slicing.

In both infalling Eddington-Finkelstein and regular coordinates
the infalling boundary condition is replaced by the condition of regularity at the horizon. There is however a difference between the coordinate systems concerning the resulting eigenvalue problem. In infalling Eddington-Finkelstein coordinates one ends up with a generalized eigenvalue problem whereas regular coordinates result in a standard eigenvalue problem. We chose the latter approach and briefly review the findings of \cite{Arean:2023ejh}.

A particular choice of regular coordinates for the black brane is 
\begin{align}\label{eq:RegularCoordinates}
\tau &=t-\left(1-\frac{1}{r}\right)+\int \frac{dr}{f(r)}\left(\frac{1}{r}\right)^2\,,\\
\rho& = 1-\frac 1 r \,.
\end{align}
in which the line element takes the form
\begin{equation}
\label{eq:General Regular metric compactified}
ds^2 = \frac{1}{(1-\rho)^2}\left(
-{f}(\rho) d\tau^2 + 
(d\vec{x})^2  +2(1-f(\rho)) d\tau d\rho +(2-f(\rho)) d\rho^2
\right) \,.
\end{equation}
Here we have set the AdS curvature scale $L=1$ and re-scaled coordinates such as to absorb the scale set by the horizon ($r_h=\pi T$). In these coordinates the boundary is at $\rho=1$ and the
horizon at $\rho=0$.

It is instructive to concentrate on a case in which we have actually exact analytic results about the spectrum of quasinormal frequencies and therefore we only consider the (transverse) gauge field at zero momentum. 
This means that we consider a gauge field of the form
$A_1(\rho,\tau,\vec x) = a(\rho) \exp(-i\omega \tau)$.
The equation of motion for this gauge field ansatz in the metric \eqref{eq:General Regular metric compactified} 
is second order in $\partial_\tau$. It can be reduced to a first order system by introducing the auxiliary field $\alpha$ and the additional equation $\alpha=-i\omega a$. The energy functional takes the form
\begin{equation}\label{eq:energynorm}
E[a,\alpha] = \int_0^1 \frac{d\rho}{1-\rho}\left(
f |\partial_\rho a|^2 + (2-f)\,|\alpha|^2
\right)\,,
\end{equation}
where we have discarded an overall volume factor stemming from the integration over the $x$ coordinates. Further we have taken into account that  $a(\rho)$ and
$\alpha(\rho)$ are Fourier modes and therefore complex valued. The equation of motion is given by
\begin{align}\label{eq:ev_transverse}
\omega
\Psi &= \mathcal{L} \Psi = i
\begin{pmatrix}
	0&1 \\
	L_1&L_2   
\end{pmatrix} \Psi\,,\\[\medskipamount]\label{eq:def_difops}
L_1 &=\frac{1}{f-2}
\left[-(1-\rho)\left(\frac{f }{1-\rho}\right)'\partial_{\rho} -f\partial_{\rho}^2 \right]\,,\\[\medskipamount]\label{eq:Eigenvalue problem Transverse Guage field 3}
L_2&=\frac{1}{f-2}\left[(1-\rho)\left(\frac{f-1}{1-\rho}\right)' +2 \left(f-1\right)\partial_{\rho}  \right]\,,
\end{align}
where $\Psi = (a,\alpha)^T$.

Quasinormal modes can now be defined as the eigenvalues of the operator $\mathcal{L}$ with  Dirichlet boundary conditions at $\rho=1$ and
regularity at the horizon $\rho=0$. The energy can be promoted to an inner product
\begin{equation}\label{eq:innerproduct}
\langle\Psi_1,\Psi_2\rangle = \int_0^1 \frac{d\rho}{1-\rho}\left[\,
f(\partial_\rho a_2)^*  (\partial_\rho a_1)+ (2-f)\,\alpha_2^* \alpha_1 \,
\right]\,,
\end{equation}
The operator $\mathcal{L}$ is self adjoint up to a boundary term with respect to this inner product
\begin{equation}\label{eq:Ldag GF}
\mathcal{L}^\dagger=\mathcal{L}+\begin{pmatrix}
	0&0\\[\medskipamount]
	0&-i\delta(\rho)
\end{pmatrix}\,,
\end{equation}
which nicely reflects the fact that dissipation stems from the boundary condition at the horizon.

We note that the inner product \eqref{eq:innerproduct} induces a norm on the space of linear operators acting on $\Psi$. This operator norm can be used to define the
$\epsilon$-pseudospectrum of $\mathcal L$ as the set in the complex $\omega$ plane where
\begin{eqnarray}
\sigma_\epsilon = \left\{ \omega \in \mathbf{C} : ||(\mathcal{L} -\omega)^{-1}|| > \frac{1}{\epsilon}\right\}\,.
\end{eqnarray}
We refer to \cite{trefethen2005spectra} for comprehensive information about the pseudospectrum.
For our purpose, the most useful interpretation is that for any operator $\delta \mathcal{L}$ of operator norm $||\delta \mathcal{L}||<\epsilon$; the spectrum of $\mathcal{L}+\delta \mathcal{L}$ lies inside $\sigma_\epsilon$

It is convenient and informative to present the pseudospectra as a contour plot in which the contour lines correspond to different values of $\epsilon$. In the case of a normal operator these contour lines are concentric circles around the eigenvalues. In particular for sufficiently small $\epsilon$ the radius of the circle is also given by $\epsilon$. This situation can be referred to as spectral stability. For non-normal operators however the contour lines are not necessarily circles. They can be much larger than circles of radius $\epsilon$ or even open lines in the complex $\omega$ plane. This indicates small perturbations can displace the eigenvalues of the operator by large amounts. 

\begin{figure}
	\includegraphics[width=.5\linewidth]{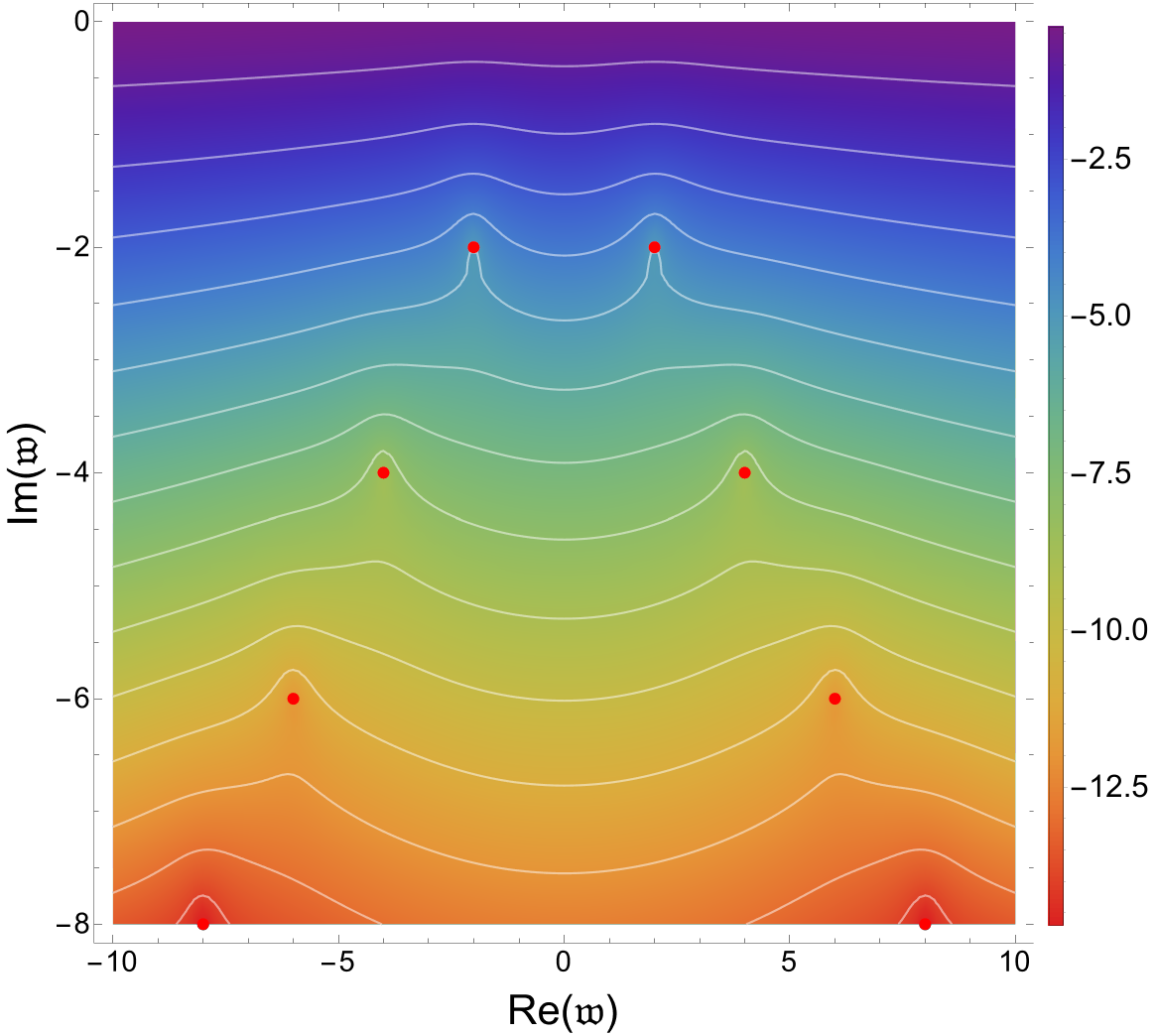}
	\caption{Pseudospectra of a vector field in the AdS black brane background. The color code indicates the values of $\log_{10}\epsilon$. }
	\label{fig:LargePseudoGFq0}
\end{figure}

Let us have a look now to the pseudospectrum in our case.
It is shown in figure \ref{fig:LargePseudoGFq0}. As one can see the contour lines are open. The colors indicate the $\epsilon$ values. Even tiny perturbations can completely destabilize the spectrum of quasinormal modes!
It is important to note that this figure has been obtained with a discretization of the differential operator $\mathcal{L}$ using pseudospectral methods at a grid size of $N=120$ points for $\rho\in[0,1]$.  It turns out that the spectral instability gets stronger as the gridsize is increased. 
In fact one can argue that the resolvent does not converge to a finite value for $N \rightarrow \infty$ \cite{Boyanov:2023qqf}. The reason is that the energy norm cannot effectively exclude the modes which are outgoing from the horizon. These behave like $a\propto \rho^{i\omega/2}$ near the horizon.
The energy norm however only demands integrability on the horizon. In fact all functions which behave like the outgoing modes with $\Im(\omega)<0$ have an integrable energy \eqref{eq:energynorm}.
Furthermore, the domain on which the operator $\mathcal{L}$ is defined contains the outgoing modes with $\Im(\omega)<-1$.
Therefore, in the continuum limit all points with $\Im(\omega)<-1$ belong to the spectrum of the operator $\mathcal{L}$. For an in-depth mathematical discussion see \cite{Warnick:2013hba,Warnick:2024usx}. We note that hydrodynamic modes for small enough momentum $k$ obey $\Im(\omega)\geq-1$ 
and thus they lie in the convergent region of the pseudospectrum in the energy norm \cite{Cownden:2023dam,Boyanov:2023qqf}.

\section{Discussion}
This result on the spectral instability of quasinormal modes is somewhat puzzling. After all we can construct the holographic Green's function exactly and it does have a discrete set of poles in the complex $\omega$ plane. In contrast the spectrum of $\mathcal{L}$ is continuous if it acts on functions with finite energy norm. 

We note that, as we have emphasized, the definition of the holographic Green's function implicitly relies on analytic continuation across the horizon. This analyticity requirement is much stronger than the requirement of existence of the energy norm. A way to circumvent this has been suggested in \cite{Warnick:2013hba} and consists in replacing the energy norm by a Sobolev norm. In physicists terms this corresponds to higher order derivative terms in the norm. Higher derivative terms up to $|\partial_\rho^n a|^2$ amount to lowering the limit for integrability to $\Im(\omega)<1-2n$.
In order to recover the exact spectrum one would of course have to take a limit with infinitely many derivatives. From the physics point of view the significance of such higher derivative terms is not clear. 

Another line of thought could be that one considers the underlying theory (being it a scalar field, a Maxwell field or the metric itself) as an effective field theory valid down to a finite cutoff length scale $\Lambda$. Then we would necessarily have some huge but finite value for $N$ determined e.g. by the criterion that the minimal distance between points of the discretization is larger than $\Lambda$. Alternatively, one could also impose the boundary conditions not directly at the horizon but slightly outside at a sort of ``stretched'' horizon \citep{Price:1986yy}.

Let us now emphasize the importance of pseudospectra in the context of holography. From the gravitational side, pseudospectra probe how much the quasinormal frequencies change if the background is slightly modified in some way (\textit{e.g.}, by a change of the geometry and/or the background value of the fields).
Consequently, in the dual quantum field theory, pseudospectra help us estimate how much the poles of the retarded Green's functions might change if the theory is slightly perturbed. In both cases, these perturbations should be understood as perturbations to the Lagrangian, leading to a change in the spectrum of excitations. Then, spectral instability suggests that holographic models might not be able to accurately capture the actual spectra of real physical systems such as quark-gluon plasma. However, valuable information, such as transient dynamics, can still be obtained by studying pseudospectra \cite{Jaramillo:2022kuv,trefethen2005spectra}.

We shall now point to additional results on quasinormal modes in Anti-de Sitter space. The pseudospectrum in infalling Eddingtion-Finkelstein coordinates has been investigated in \citep{Cownden:2023dam}. One of the main findings was that in certain cases the pseudospectrum can significantly reach up into the upper halfplane giving rise to possible transient behaviour. Structural aspects of the pseudospectrum of quasinormal modes for AdS black holes have been pointed out and further investigated in \cite{Boyanov:2023qqf}. In particular the results in infalling Eddington-Finkelstein and regular coordinates have been contrasted. The dependence of pseudospectra on the choice of coordinates still needs further investigation. The properties of the pseudospectrum of black hole metrics have also been shown to give rise to transient behaviour for which a sum of $M$ quasinormal modes can be long lived of order $\log(M)$ in \cite{Carballo:2024kbk}. The stability of complex linear momenta ($\mathbb{C}$LMs) in Anti-de Sitter was studied in \cite{Garcia-Farina:2024pdd}. Remarkably the pseudospectrum of $\mathbb{C}$LMs was observed to be convergent allowing for quantitative results.

In this paper we have reviewed the holographic perspective on quasinormal modes and quasinormal frequencies of AdS black holes.
In this context, pseudospectrum analysis offers an invaluable tool for assessing the stability as well as investigating the existence of transient dynamics. Numerically-computed pseudospectra do not converge in the energy norm because outgoing modes can still have finite energy. We believe that the lack of convergence is not a flaw of the construction but rather a fundamental feature which needs to be addressed using a physics-motivated regulator.

\section*{Funding}
This work is supported through the grants CEX2020-001007-S and PID2021-123017NB-100, PID2021-127726NB-I00 funded by MCIN/AEI/10.13039/501100011033 and by ERDF ``A way of making Europe''. 
The work of D.G.F. is supported by FPI grant PRE2022-101810.

\section*{Acknowledgments}
We thank V. Boyanov and especially J.L. Jaramillo for numerous insightful discussions on the properties of quasinormal modes and their pseudospectra.

\bibliographystyle{apsrev4-1}
\bibliography{test}

\end{document}